%% file: main.tex
\documentclass[AMA,STIX1COL]{WileyNJD-v2}
\usepackage{float}

\articletype{Research Article}%

\received{26 April 2016}
\revised{6 June 2016}
\accepted{6 June 2016}

\usepackage{blindtext}
\usepackage{subfiles} 
\usepackage{algorithm}
\usepackage{algpseudocode}
\usepackage{hyperref}

\usepackage{comment}

\begin{document}
\title{Enrollment Forecast for Clinical Trials at the Planning Phase with Study-Level Historical Data}

\author[1]{Mengjia Yu}

\author[1]{Sheng Zhong*}

\author[1]{Yunzhao Xing}

\author[1]{Li Wang*}

\authormark{Yu \textsc{et al.}}

\address{\orgdiv{Statistical Innovation Group, Data and Statistical Sciences}, \orgname{AbbVie Inc.}, \orgaddress{\state{Illinois}, \country{United States}}}

\corres{Li Wang,
\email{wangleelee@gmail.com}\\
Sheng Zhong, \email{zhongever@gmail.com}
}

\presentaddress{1 N Waukegan Rd, North Chicago, IL, 60064}

\abstract[Abstract]{
Given progressive developments and demands on clinical trials, accurate enrollment timeline forecasting is increasingly crucial for both strategic decision-making and trial execution excellence. 
Naïve approach assumes flat rates on enrollment using average of historical data, while traditional statistical approach applies simple Poisson-Gamma model using time-invariant rates for site activation and subject recruitment. Both of them are lack of non-trivial factors such as time and location.
We propose a novel two-segment statistical approach based on Quasi-Poisson regression for subject accrual rate and Poisson process for subject enrollment and site activation.
The input study-level data is publicly accessible and it can be integrated with historical study data from user's organization to prospectively predict enrollment timeline. The new framework is neat and accurate compared to preceding works. We validate the performance of our proposed enrollment model and compare the results with other frameworks on 7 curated studies. 
}

\maketitle

\section{INTRODUCTION}

\subfile{sections/introduction}

\section{INPUT DATA AND FORECAST WORKFLOW}

\subfile{sections/dataworkflow}

\section{METHODOLOGY DETAILS}

\subfile{sections/methods}

\section{PERFORMANCE EVALUATION}

\subfile{sections/performance}

\section{DISCUSSION}

\subfile{sections/discussion}

\section*{Disclosure}
This manuscript was sponsored by AbbVie. AbbVie contributed to the design, research, and interpretation of data, writing, reviewing, and approved the content.   All authors are employees of AbbVie Inc. and may own AbbVie stock.

\section*{Data availability statement}

The authors elect to not share data.


\newpage
\bibliography{wileyNJD-AMA}
\clearpage

\end{document}

%% file: sections/introduction.tex
In clinical development, forecast of trial duration especially at the portfolio planning stage is extremely important. It facilitates the informed decision making for senior management and provides an initial idea of the potential spend on time and cost for any new trial proposal. An advanced predictive modeling providing accurate enough enrollment forecast across all portfolios is in starving demand. There are several quantitative approaches developed in statistical literature.  

A naive approach is based on Patients (subjects) per Site per Month, often abbreviated as $psm$, which is defined as 
$psm = \textit{number of enrolled patients or subjects} / (\textit{number of sites} \times \textit{enrollment time})$.
Assuming that every site in a trial will have such a constant enrollment rate over time, the projected enrollment duration for a planned trial is simply 
$\textit{planned sample size}/(\textit{number of sites proposed} \times psm)$,
where the $psm$ rate in the denominator can be directly taken from the average rate in historical studies or user past experience.  This approach is simple and empirical. It relies on strong assumption on the constant enrollment rate for all sites overtime, which is not usually the case in reality. With often unsatisfactory predictive performance from this naive approach, a more sophisticated statistics modeling is needed to fill the gap.

There are various statistical approaches proposed in the literature to model and predict patient accrual. The papers \cite{anisimov2016discussion, HEITJAN201526} are provoking systematic literature review for patient accrual models. 
Bagiella and Heitjan \cite{https://doi.org/10.1002/sim.843} proposed to use a homogenous Poisson process to model the patient recruitment. 
Anisimov and Fedorov \cite{https://doi.org/10.1002/sim.2956} improved the recruitment
model by applying a Poisson-Gamma mixture model to handle the variation in recruitment rate across multiple centers.
As pointed out by succeeding works \cite{anisimov2020modern, anisimov2011statistical, anisimov2007recruitment}, such type of methods is mainly random effects models. The intrinsic goal is to capture the heterogeneity in enrollment
rates across different centers whereas the enrollment pattern within each center is described by a homogeneous
Poisson process with gamma distributed rate. 
Lan, Tang, and Heitjan \cite{https://doi.org/10.1002/sim.8036} first proposed a time-varying rate function that allows modeling
the time decay trend in recruitment while taking site initiation into consideration. 
Deng, Zhang, and Long \cite{deng2017bayesian} investigated a Bayesian approach using a non-homogeneous Poisson process where region-specific accrual is accounted in their framework. 
Zhang and Long \cite{zhang2010stochastic,zhang2012joint} employed a non-homogeneous Poisson process to model patient accrual where the underlying accrual rates are allowed to change over time. 
Wang et al. \cite{wang2022real} defined time to endpoint maturation framework and linked the concept to key milestone dates in clinical trials. They proposed a simulation based non-homogeneous Poisson process with a normal kernel enrollment rate which can capture the up-and-down enrollment trend in reality and provided improved prediction performance in both simulated and real study enrollment data.   

Motivated by 
Wang et al. \cite{wang2022real} and others work, 
Zhong et al. \cite{zhong2022sitelevel} proposed a novel statistical framework based on generalized linear mixed-effects model (GLMM) and the use of non-homogeneous Poisson processes through Bayesian hierarchical modeling framework to predict trial duration at the portfolio planning stage. It utilizes site level enrollment information from proprietary data to predict trial duration more accurately based a set of pre-selected validation studies. It also shows that their modeling and simulation approach calibrates the data variability appropriately and gives correct coverage rates for prediction intervals of various nominal levels.

One of the challenges for 
Zhong et al. \cite{zhong2022sitelevel} is the availability and the size of the site level enrollment information. Some site-level data is proprietary and may be not publicly available, where the number of studies from that data is just limited to several sponsor companies who contributed to the data set. To make the prediction algorithm beneficial to broader users in pharmaceutical industry, we proposed a new advanced statistical modeling algorithm utilizing publicly available study-level information that can be generalized from \url{ClinicalTrials.gov}. This data source is free and it contained extensive number of studies compared to proprietary data. 
The study-level information from \url{ClinicalTrials.gov} together with historical data within users organization can bring prospective evaluations on subjects from new sites and for new indications. 
Our methodology includes different homogeneous Poisson process for subject enrollment and site activation. Quasi-Poisson regression and Monte Carlo sampling are proposed to estimate and simulate subject enrollment, and a linear (fixed) approach and a perturbed approach are proposed to model site activation. 

The rest of this paper is organized as follows. The input data and entire forecast workflow are described in Section 2. In Section 3, we derive the details of methodology. Section 4 provides the prediction performance based on real case studies. Discussions are given in Section 5.

%% file: sections/dataworkflow.tex
\begin{figure}[h]
\centering
\caption{Two Sequential Segments of Enrollment Framework}
\label{framework_figure}
\includegraphics[width=16cm]{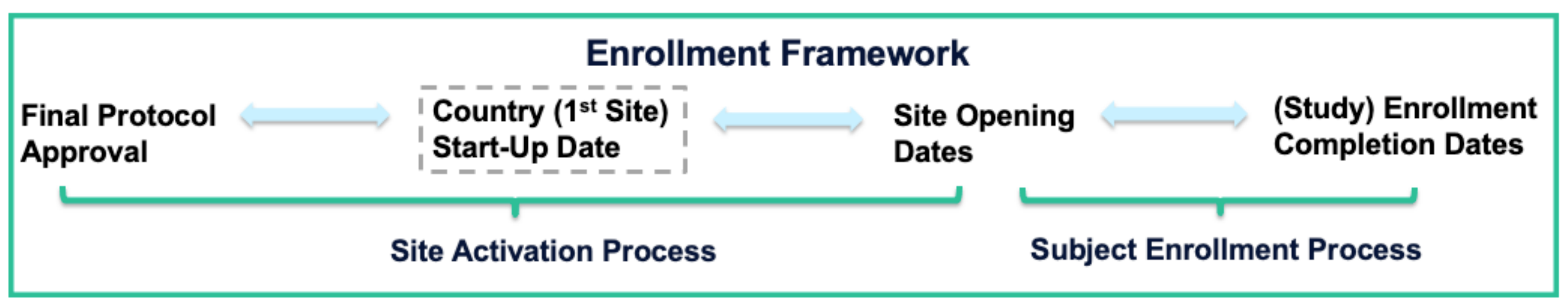}
\end{figure}

\label{sec:data_flow}
The whole enrollment procedure in clinical trials is a complex process with multiple steps. Previous research usually concentrated on the key part of subject enrollment \cite{anisimov2020modern, anisimov2011statistical, anisimov2007recruitment}. Nevertheless, subject enrollment is just one factor in the clinical trial enrollment procedure. In real-world practice, the country/site preparation step (such as contracting and training investigators, preparing doses, etc) that is prior to subject enrollment can also be a compelling impact on the enrollment timeline. This paper proposes a comprehensive enrollment framework based on study-level historical trial data, which covers site activation and the subject enrollment processes. Models with proper parameter estimation and simulation will be developed (depending on historical data availability) to cover each part of the enrollment framework. Unless stated otherwise, the entire enrollment process in a study, as illustrated in Figure~\ref{framework_figure}, is referred to as two sequential segments from the date of final protocol approval to the date of subject enrollment completion. 

The first segment of the enrollment framework is the site activation process, defined as the time between the study start date and the opening date of each site. We assume all the selected sites and corresponding countries in a clinical trial will share the same study start date defined as the final protocol approval date. 
In addition, we assume sites are indistinguishable within each country and allow each country's first site activation date to be modeled separately from the others. The first site activation is also referred as country start-up, a country-level procedure involving country approval and preparation activities such as country/region regulatory approval and IRB approval process. In real world, it can be significantly affected by a couple of factors, including but not limited to sponsors, countries/regions, therapeutic areas, and phases. Therefore, dedicated integrated data from multiple sources are required to predict the country's start-up and site activation. From a modeling perspective, data availability is also a critical factor, since final protocol approval dates and site activation dates are typically unavailable in the most common public databases. Our proposed model will rely on an internally available data set to develop a forecasting process for site activation.

The second part of the enrollment framework is the subject enrollment process. It is a country-specific procedure that starts from the country start-up date to the enrollment completion date of the entire clinical trial. All the sites in the same country start to recruit patients after the country's start-up date. No particular site start-up patterns are considered due to the limitation of study-level historical data. The enrollment rate of historical trials will be adjusted based on the simulated country start-up pattern to avoid bias.

%% file: sections/methods.tex
\subsection{General Description}

The ultimate goal of our proposed framework is to predict enrollment timeline of a study at planning stage. With given number of sites in each country, a forecast of enrollment duration can be accomplished if we can model the opening time of sites and the corresponding subject recruitment pattern of each site.
In our framework, we assume the number of sites in each country is pre-specified, and the models will be built under given therapeutic area. 
We do not make any assumption on specific recruitment criterion or in-trial observation to be available. Instead, historical data, both internally and externally, plays a role in parameter estimation in each step. 

Our model consists of two parts: subject enrollment in each site and site activation in each country. For the subject enrollment model, a homogeneous Poisson Process is applied, where the rate is assumed to be the same across all sites in all countries and can be estimated from historical studies using quasi-Poisson regression. Monte Carlo sampling of subject accrual rate is also considered.  
For the site activation model, another homogeneous Poisson Process is applied, where the rate is estimated and simulated from historical data that are internally available within the author's organization. A simplified linear (fixed) approach and a perturbed approach are proposed. We will describe the model details in the next and provide a summarized modelling procedure at the end of this section.

\subsection{Poisson Process for Subject Enrollment Model}
Let $N_{ev}(u_1,u_2)$ be the number of subjects enrolled between time points $u_1$ and $u_2$. Suppose there are $n_{cntry}$ countries, where for the $i$-th country there are $n_{cntry,i}$ sites open for recruitment of subjects. Let $N_{ev,i,j}(u_1,u_2)$ be the number of subjects enrolled from the $j$-th site in the $i$-th country. Then, we assume the subject enrollment in each site is an independent Poisson Process
\[
N_{ev}(u_1,u_2) = \sum_{i=1}^{n_{cntry}} \sum_{j=1}^{n_{cntry,i}} N_{ev,i,j}(u_1,u_2),
\] 
\[
N_{ev,i,j}(u_1,u_2) \overset{\mathrm{independent}}{\sim} Poisson \left( \int_{u_1}^{u_2} \lambda_{ij}(\nu)  d \nu \right) ,
\]
where $\lambda_{ij}(=\lambda_i)$ is the subject accrual rate that are the same across all sites in a county.

In our framework, we simply use linear subject accrual rate $\lambda_i(\nu) = \mu$ that are the same across all countries. A common example for the concept of $\mu$ in clinical trials is the Patient per Site per Month ($psm$).  Given the site activation (opening) time point $u_{i,j}, i=1,\dots,n_{cntry}, j=1,\dots,n_{cntry,i}$, the rate parameter in Poisson Process can be simplified as
\[
\int_{u_1}^{u_2} \lambda_{ij}(\nu)  d \nu =  (u_2\vee u_{i,j}- u_1\vee u_{i,j}) \mu,
\]
where $a \vee b$ is the maximum between $a$ and $b$. Therefore, the subject enrollment model becomes
\[
N_{ev}(u_1,u_2) \sim Poisson \left( \sum_{i=1}^{n_{cntry}}  \sum_{j=1}^{n_{cntry,i}} \{u_2\vee u_{i,j}- u_1\vee u_{i,j}\} \mu  \right).
\]
Here, we assume the list of countries and the number of sites in each given country are usually given by the study team. If they are pending estimated, a country-site recommendation algorithm will be discussed in Section~\ref{sec:discussion}. With the number of sites in each countriy determined, we only need to know $\mu$ and $u_{i,j}$ to generate the subject enrollment pattern.

\subsection{Quasi-Poisson Estimation of Subject Accrual Rate $\mu$}
Since we assume the study in plan is for a new drug or a new indication within our organization, a better source for subject accrual rate estimation is the global data from Citeline (ClinicalTrials.gov) rather than the internal Clinical Trial Management System (CTMS). To circumvent the caveat that Citeline does not include site-level information, the subject accrual rate $\mu$ must be modeled directly.
Suppose there are $s=1, \dots, S$ historical studies to be used in this estimation step, where the $S$ studies are filtered from Citeline under certain criteria that are subject to specific requirement for the study to forecast. For example, the criteria can include study phase, therapeutic area, indication and etc.
For each study, denote $T_{i,j}^{(s)}$ as the enrollment duration for Site $j$ in Country $i$, $X_s$ as the number of accrual subjects and $d_s$ as the offset of combined effect from number of sites and enrollment duration. The setup of quasi-Poisson is
\[
X_s \sim quasi-Poisson (\lambda_s),  \quad E(X_s) = \phi \ var(X_s) \quad \text{and} \quad \log \lambda_s = \log d_s + \mu.
\]
Here, we allow an over-dispersion effect through the dispersion parameter $\phi$ such that $E(X_s) = \phi \ var(X_s)$ and 
\[
d_s = \sum_{i=1}^{n_{cntry}^{(s)}} \sum_{j=1}^{n_{cntry,i}^{(s)}} T_{i,j}^{(s)}.
\]

\begin{figure}[h]
\centering
\caption{Offset adjustment for subject accrual estimation in quasi-Poisson estimation.}
\label{fig:offset_adjust}
\includegraphics[width=7cm]{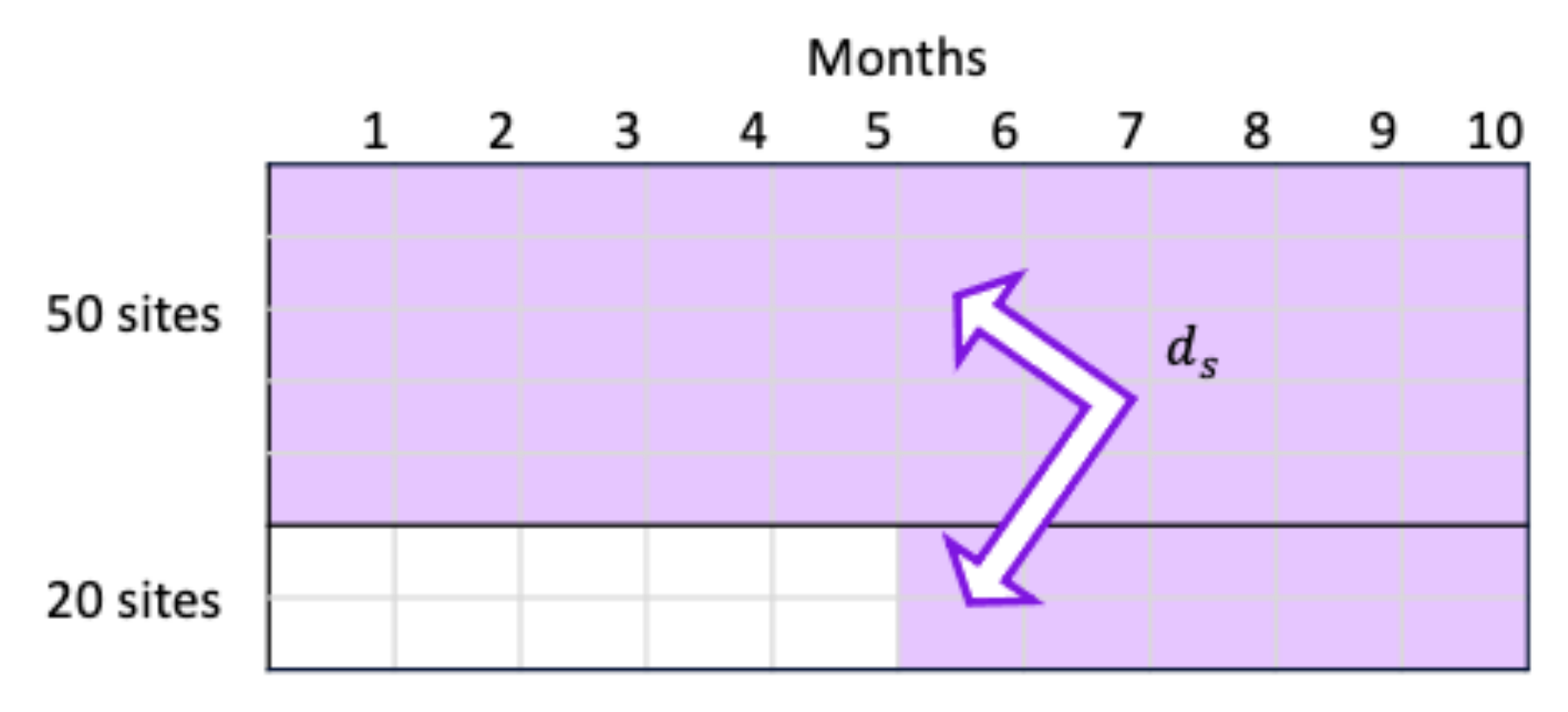}
\end{figure}

The offset $d_s$ serves as a calibration in modeling the rate $\mu$ using count data. In most cases, the widely-used $psm$ simply reflects the total number of subject divided by the total number of sites and entire study duration. However, due to operational considerations and other reasons, countries can start up in different pace. Hence, the $psm$ before any adjustment can be slower and leads to a forecast departing far from reality. We propose to adjust the $d_s$, which should be in the magnitude of "site*month", by integrating the opening time of each site, approximately or exactly upon data availability. For example, suppose the enrollment of study lasted 10 months, where Country~1 opened 50 sites at Day 1 and Country~2 opened 20 sites at Month 6. Then, $d_s$ for this historical trial becomes 600 site*month as illustrated in Figure~\ref{fig:offset_adjust}. Ideally, when each site opening time is documented and available to public, it is better to calculate the $d_s$ from such granule information. However, country start-up times sometimes are even unknown. Under such circumstances, we suggest applying the idea in Section~\ref{sec:fixed_estimate} to estimate $T_{i,j}^{(s)}$.

In real application, over-dispersion arises naturally when a model does not contain a parameter for modeling the variance directly. Some common cases in generalized linear models include multinomial distribution based models like logistic regression where the variance is a function of the mean, and count data models like Poisson regression where variance equals mean. To better incorporate variability in prediction of parameter distribution, Monte Carlo simulation is applied to draw the subject accrual rate given model estimate $\hat{\mu}$ and the corresponding standard deviation $\sigma_{\mu}$. For example, one can obtain a sample of $\mu_b, b=1,\dots,B$ from normal distribution $N(\hat{\mu},\sigma_{\mu}^2)$ in practice.

\begin{remark}
Alternatively, random effect of $\mu$ may be introduced to the Poisson model to account for over-dispersion effect, where 
\[
X_s \sim Poisson (\lambda_s), \quad \log \lambda_s = \log d_s + \mu + \tau_s, \quad \tau_s \overset{\mathrm{i.i.d.}}{\sim} N(0,\sigma_\tau^2).
\]
Through this model, the average enrollment rate from different studies can be conceptualized as a realization of population level effect for a type of indication or a category of therapeutic area. Different from the fixed quasi-Poisson model where we can only estimate the mean of accrual rate for the $S$ studies that share a common variance, we allocate the variance to broader (but hidden) group level effect such that the inference can be extended to similar collection of trials. However, random effect model usually requires moderate number of observations. 
Due to the limits of real data availability, we do not consider random effect model for now. The random effect of indication or therapeutic area can be left to future exploration. 
\end{remark}

\subsection{Poisson Process for Site Activation $u_{i,j}$ and the linear (fixed) projection} \label{sec:fixed_estimate}
Denote $N_{site,i}$ as the number of sites opened in Country $i$ and $\beta_i$ as the corresponding site opening rate. Recall $u_{i,j}$ as the $j$-th site activation time in Country $i$. We assume the sites are non-distinguishable and their activation time points follow the following Poisson Process
\[
N_{site,i}(u_2) - N_{site,i}(u_1) \sim Poisson \left( (u_2 \vee t_i - u_1\vee t_i) \beta_i \right),
\]
where $t_i = u_{i,1}$ is the first site initiation time across Country $i$, i.e., the start-up time of Country $i$. Denote $w_k$ as the whole enrollment duration for Study $k$.
Consequently, 
\[
u_{i,j} = \inf_u \{N_{site,i}(u)=j\}, \quad T_{i,j}^{(k)} = w_k - u_{i,j}.
\]

In clinical trials, from protocol approval to the first site opening, the preparation period including but not limited to building database, developing Trial Master File (TMF), training Principal Investigators, generating subject randomization schedule and preparing investigational kits can take months to a year or even long. Hence, we must estimate $t_i$ separately from $u_{i,j}$ or $T_{i,j}^{(k)}$ for better accuracy. Here, we rely on internal CTMS data rather than Citeline data to extract anchor dates due to data availability.
Our framework first uses the median of historical site initiation time within each country as 
\[
\hat{t}_i = median_k \{t_i^{(k)} | \textnormal{$t_i^{(k)}$ is the first site initiation time of Country $i$ for all available studies $k=1,\dots$}\}.    
\]
As all site activation times are available for each study in CTMS, the duration between the first site activation and the last site activation can also be found in summary. Then, the median of their ratios (i.e., the total amount of sites divided by the duration of site opening) can serve as the estimate of $\hat{\beta}_i$. 
To further simplify the Poisson Process, linear (fixed) projection can be used to approximate $N_{site,i}(u)$ and $T_{i,j}^{(k)}$ so that the time between any consecutive site openings are equal. Specifically, one can simply estimate $\hat{u}_{i,j} = \hat{t}_i + (j-1)\hat{\beta}_i$. 

Similarly,  this idea can also apply to the $T_{i,j}^{(s)}$ 
in calibrating $d_s$ from the previous section, if deemed necessary. Let $\hat{T}_{i,j}^{(s)} = w_s - \hat{u}_{i,j}$, where $w_s$ is the whole enrollment duration for Study $s$, $s=1,\dots, S$ and $\hat{u}_{i,j}$ is the activation time for the $j$-th site in Country $i$. The subscript runs within $i=1,\dots,n_{cntry}, j=1,\dots,n_{cntry,i}^{(s)}$, where $n_{cntry,i}^{(s)}$ is the total number of sites activated in Country $i$ for historical Study $s$. In application, it is common to have $w_s$ available, but $n_{cntry,i}^{(s)}$ from Citeline may be unreliable sometimes. Hence, one may further approximate $d_s$ by making $\hat{T}_{i,j}^{(s)} = w_s$ so the summation in $d_s$ reduces to the product of trial duration and total number of sites. In our numeric study, we simply plugin the estimations from internal CTMS data.

\subsection{Country-level Perturbation for Site Opening Rate $\beta_i$ and Site Initiation $t_i$} \label{sec:perturbation_estimate}
In model development, when data and model are consistent, it is always believed that a greater amount of data points can depict a more accurate and convincing story. However, when it comes to forecasting models in practice, a narrow prediction window (prediction interval) can be restrictive to guide team on study preparation or budget calculation. Another criticism is that future projection may easily shift from historical record subtly.  Therefore, we recommend introducing some randomness to avoid such spurious precision issue by performing parameter perturbation based on historical country-level data. 

For example, we can bootstrap site activation time (or site opening rate) of Country $i$ as estimates of $t_i$ (or $\beta_i$) instead of using the fixed median. We can also draw the paired samples of $t_i$ and $\beta_i$ for each country to maintain their correlation, which is commonly positive. In fact, site preparation, such as contracting and training, always starts way ahead of protocol approval in clinical trials. Delayed activation of one site (in a country) may not heavily impact other sites to be operationally ready soon, thus a faster opening pattern of $\beta_i$ usually accompany prolonged country start-up $t_i$. Besides, if a trial defers site activation, the study team is likely to urge subsequent sites to open and enroll patients to catch up timeline. As a result, the paired bootstrapping of site initiation time and site opening rate is more desirable.

\subsection{Summary of Modelling Procedure}
The Algorithm~\ref{alg:cap} provides pseudo code for our study-level enrollment prediction model.

\begin{algorithm}
\caption{Algorithm of study-level enrollment model}\label{alg:cap}
\begin{algorithmic}
\Require $u_1 \gets 0$  \Comment{E.g., The time zero can be protocol approval date.}
\Require Therapeutic area, list of countries, and etc. to filter historical data
\Ensure Anticipated amounts of sites $n_{cntry,i}$ for $i=1,\dots,n_{cntry}$ and enrolled subjects $n_{ev}$ are given.

    \For{$b=1, b++, b\leq B$}
    \State \textbf{Site Activation:}
        \State Estimate $\hat{t_i}$ for $i=1,\dots,n_{cntry}$
        \State Estimate $\hat{\beta_i}$ for $i=1,\dots,n_{cntry}$
        \State $\hat{u}_{i,j} \gets \hat{t}_i + (j-1)\hat{\beta}_i$ and $T_{i,j}^{(s)} = w_s - u_{i,j}$ for $j = 1, \dots, n_{cntry,i}$
        \State Obtain offsets for historical studies $\hat{d}_s$ for $s=1,\dots, S$ 
    \State \textbf{Subject Enrollment:}
        \State Perform quasi-Poisson model and estimate $\hat{\lambda}_s$ for $s=1,\dots, S$ 
        \State Sample $N_{ev,i,j}(u_1,u_2)$ independently and sum together
    \EndFor
\end{algorithmic}
\end{algorithm}

%% file: sections/performance.tex
\subsection{General Consideration and Data Description}

We will apply our study-level prediction model to 7 recently completed studies within our organization and compared the prediction to their actual enrollment duration, which is the period between LSFD (last subject first dose) and
the protocol approval. 

The data for estimating site activation are internally available CTMS data within the author's organization.
In the subject enrollment modelling process, we filtered out interventional studies that are within the same therapeutic area (oncology, neuroscience, immunology, and general medicine) and the same study phases (Phase 2 and/or 3) as the target study to forecast. 
The data for modelling subject enrollment contain both internal and external data that are available on Citeline. The information include therapeutic area, 
study phase,
study duration,
the total number of subjects, 
the total number of sites, 
disease indication 
and the patient population. 
Here, the selected studies must be interventional, Phase 2 and/or 3, completed and with the same therapeutic area, disease indication and patient population as the study to be tested. Our 7 candidate studies for performance testing purpose are curated and prepared by an independent data management team to ensure data objectivity and integrity. The study numbers are masked due to confidential reason.

We also compare the performance of our proposed modeling framework to a previous internally-developed
modeling pipelines, based on the classical Poisson-Gamma models and a new site-level modeller \cite{zhong2022sitelevel}. 

\subsection{Model Performance}
\label{sec:model_performance}
\subsubsection{The approach with linear (fixed) estimates of site opening}
We first apply our model and use the linear (fixed) estimation of $t_{i}$ and $\beta_{i}$ as described in Section~\ref{sec:fixed_estimate}. 
In Table~\ref{tab:table1}, we report the actual enrollment duration followed by the predicted value and the corresponding prediction error. We also provide the 95\% Prediction Interval (PI) and the coverage status within 95\% PI, +/- 1 month, +/- 2 months and +/- 3 months. In order to better investigate the source of modelling validity, we also list the actual FSFD (first subject first dose) and the predicted FSFD together with the corresponding 95\% PI to reflect the prediction accuracy on the site initiation $t_i$. 

From the predictions of enrollment we can find that the linear approach gives more accurate estimation on Studies 2-7 compared to that of Study 1.
The 95\% PI covers 4 studies, but their PI windows are also wide and may be lack of usability. Hence, we should look at the fixed-window coverage as well.
The 3-month coverage overall shows that our prediction of enrollment performs well. When we compare the actual FSFD (first subject first dose) with the predicted FSFD, the forecasts from the linear (fixed) approach are generally aggressive, which indicates a negative bias from the site activation portion. 

\begin{table}[h]
\begin{tabular}{p{0.04\linewidth}|p{0.065\linewidth}p{0.065\linewidth}p{0.065\linewidth}p{0.072\linewidth}p{0.056\linewidth}p{0.056\linewidth}p{0.056\linewidth}p{0.056\linewidth}|p{0.055\linewidth}p{0.055\linewidth}p{0.06\linewidth}}
\hline
Study & Actual Enrollment (mo) & Predicted Enrollment  (mo) & Prediction Error (mo) & 95\%PI (mo)    & Within 95\%PI & Within +/-1 month & Within +/-2 month & Within +/-3 month & Actual FSFD (mo) & Predicted FSFD (mo) & 95\%PI (mo)   \\ \hline
1     & 26.4                            & 19.3                               & -7.0             & (15.5,24.0) & NO       & NO         & NO         & NO         & 5.3          & 4.6                   & (4.1,6.2) \\
2     & 17.5                            & 18.9                               & 1.4              & (15.8,22.4) & YES        & NO         & YES          & YES          & 7.9          & 4.2                   & (3.1,5.3) \\
3     & 11.3                            & 9.2                                & -2.1             & (8.1,10.3)  & NO       & NO         & NO         & YES          & 6.3          & 3.1                   & (3.0,3.8) \\
4     & 10.0                            & 8.7                                & -1.3             & (7.7,9.9)   & NO       & NO         & YES          & YES          & 7.0          & 3.2                   & (3.0,4.0) \\
5     & 21.1                            & 20.1                               & -1.0             & (16.0,26.5) & YES        & NO         & YES          & YES          & 8.4          & 4.1                   & (3.1,5.2) \\
6     & 16.1                            & 19.0                               & 2.9              & (14.9,24.1) & YES        & NO         & NO         & YES          & 6.0          & 5.4                   & (4.1,6.7) \\
7     & 10.8                            & 11.2                               & 0.3              & (8.0,18.1)  & YES        & YES          & YES          & YES          & 5.4          & 4.5                   & (4.0,5.6) \\ \hline
\end{tabular}
\caption{\label{tab:table1} Study-level prediction performance using the linear (fixed) approach.}
\end{table}

\subsubsection{The approach with perturbation-based approaches of site opening}
Next we apply the model using perturbation-based simulation introduced in Section ~\ref{sec:perturbation_estimate}. A similar display of prediction on the 7 studies is provided in Table~\ref{tab:table2}.

The predictions of enrollment shows that our perturbation-based approach generally performs well on Studies 2-7 compared to Study~1. The 95\% PI coverage and 2-month coverage show that our prediction captures variability from data meanwhile providing accurate estimates. 
When we compare the actual FSFD with the predicted FSFD, the perturbation-based predictions are still aggressive, but the corresponding 95\% PIs are wider compared to the linear (fixed) approach.   

\begin{table}[h]
\begin{tabular}{p{0.04\linewidth}|p{0.065\linewidth}p{0.065\linewidth}p{0.065\linewidth}p{0.072\linewidth}p{0.056\linewidth}p{0.056\linewidth}p{0.056\linewidth}p{0.056\linewidth}|p{0.055\linewidth}p{0.055\linewidth}p{0.065\linewidth}}
\hline
Study & Actual Enrollment  (mo) & Predicted Enrollment  (mo) & Prediction Error (mo) & 95\%PI (mo)    & Within 95\%PI & Within +/-1 month & Within +/-2 month & Within +/-3 month & Actual FSFD (mo) & Predicted FSFD (mo) & 95\%PI (mo)   \\ \hline
1     & 26.4                            & 18.5                               & -7.9             & (14.7,24.9) & NO       & NO         & NO         & NO         & 5.3          & 3.3                   & (1.7,6.9) \\
2     & 17.5                            & 18.8                               & 1.3              & (15.3,23.1) & YES        & NO         & YES          & YES          & 7.9          & 4.1                   & (1.3,7.3) \\
3     & 11.3                            & 9.7                                & -1.5             & (6.7,15.6)  & YES        & NO         & YES          & YES          & 6.3          & 3.3                   & (1.1,9.5) \\
4     & 10.0                            & 9.4                                & -0.6             & (6.1,15.9)  & YES        & YES          & YES          & YES          & 7.0          & 4.1                   & (1.2,10.5) \\
5     & 21.1                            & 20.1                               & -1.1             & (16.0,26.5) & YES        & NO         & YES          & YES          & 8.4          & 4.2                   & (1.2,7.2) \\
6     & 16.1                            & 19.3                               & 3.2              & (15.4,25.2) & YES        & NO         & NO         & NO         & 6.0          & 5.5                   & (4.1,9.4) \\
7     & 10.8                            & 10.3                               & -0.5             & (6.0,19.7)  & YES        & YES          & YES          & YES          & 5.4          & 3.6                   & (1.4,7.7)
\\ \hline
\end{tabular}
\caption{\label{tab:table2} Study-level prediction performance using perturbed approach.}
\end{table}

Table~\ref{tab:table3} provides a comparison between the two approaches in a summary level. The median length of 95\% PI for the perturbed approach is wider, and its prediction is more accurate as the median prediction error is smaller. They demonstrate the improvement of the perturbed approach by capturing the positive correlation between site initiation time and site opening rate. Besides, except for the 3-month coverage (of Study 6), the perturbation approach has better coverage rates than the fixed approach. We conclude that the two approaches are comparable, but the perturbation model turns to provide slightly better point prediction and wider PI on enrollment. 

In practice, a slightly narrower prediction interval that is about +/-3 month may be preferable to provide guidance for operational functions. One can use a lower level for PI calculation. For example, the median lengths of 70\% and 80\% PIs from the perturbed estimation approach are 5.3 and 6.9, respectively, while 5 out of 7 studies are covered by their own 70\% PIs as well as 80\% PIs. Hence, from application perspective, the trade-off between PI coverage and PI length need to be evaluated case by case.

\begin{table}[h]
\centering
\begin{tabular}{c|c|c|cccc}
\hline
                   & 95\%PI length                    & Prediction Error                 & \multicolumn{4}{c}{Coverage Rate}    \\
                   & \multicolumn{1}{c|}{median (mo)} & \multicolumn{1}{c|}{median (mo)} & 95\%PI & +/-1 month & +/-2 month & +/-3 month \\ \hline
Fixed estimation    & 8.46                             & -1.02                            & 57\%   & 14\%    & 57\%    & 86\%    \\
Perturbed estimation & 9.78                             & -0.60                            & 86\%   & 29\%    & 71\%    & 71\%    \\ \hline
\end{tabular}
\caption{\label{tab:table3} Comparison of prediction performance between the two proposed approaches.}
\end{table}

\subsection{Comparison with existing models}

We compare our proposed perturbation approach with the site-level model and the traditional model described in Zhong et al. \cite{zhong2022sitelevel}. 
To ensure fairness, we will focus on the same 7 studies that are Studies 6, 13, 14, 15, 16, 24 and 25 in Zhong et al. \cite{zhong2022sitelevel}.
Table~\ref{tab:table4} lists the means and medians of 95\% PI lengths and absolute prediction errors in month (mo), which are followed by the coverage rates of 95\%PI, +/-1 month, +/-2 month and +/-3 month. Detailed predictions of the Site-level model and the traditional model can be found in Table~\ref{tab:table5} and Table~\ref{tab:table6}, respectively.

\begin{table}[h]
\begin{tabular}{c|c|c|c|c|cccc}
\hline
                   & \multicolumn{2}{c|}{95\%PI length (mo)} & \multicolumn{2}{c|}{\begin{tabular}[c]{@{}c@{}}Absolute Prediction \\ Error (mo)\end{tabular}} & \multicolumn{4}{c}{Coverage Rate}    \\\cline{2-9} 
                   & median              & mean              & median      & mean  & 95\%PI & +/-1 month & +/-2 month & +/-3 month \\ \hline
Perturbation approach & 9.8                 & 10.1              & 1.3                       & 2.3                     & 86\%   & 29\%    & 71\%    & 71\%    \\
Site-level model   & 17.6                & Inf               & 1.1                       & 4.3                     & 100\%   & 29\%    & 57\%    & 71\%    \\
Traditional model  & 1.6                 & 2.4               & 4                         & 5.3                     & 29\%   & 29\%    & 29\%    & 43\%    \\ \hline
\end{tabular}
\caption{\label{tab:table4} Comparison among the proposed Study-level model, the Site-level model and the traditional model in Zhong et al. \cite{zhong2022sitelevel}.}
\end{table}

From Table~\ref{tab:table4}, we can draw the following observation.
The traditional model has the narrowest 95\% PI length, but its prediction deviation (in terms of absolute prediction error) is the largest and the coverage rates in terms of all 4 criteria are the lowest. So this model tends to provide biased predictions and fails to capture the variability from data.
The site-level model provides the widest 95\% PI so that the corresponding coverage rate reaches 100\%. However, as the authors mentioned, the site-level model take into account many sources of randomness, which is accumulated from the three-segment framework consisting of country start-up, site initiation and mixed-effect Poisson regression for subject enrollment, as well as Monte-Carlo simulation of recruitment. Hence, a particular simulation of site-level model may accidentally fall into a stagnant side and cannot enroll enough samples within a long enough period (2000 days in this case). So it leads to the infinity of the upper PI bound for Study 6. But the accuracy with respect to median absolute prediction error and 1$\sim$3 month coverage rates indicate that the site-level model in general can provide good point estimation.
Our proposed model using the perturbation approach can generate moderate 95\% PI while its absolute prediction error is comparable with the site-level model. Our 2$\sim$3 month coverage rates are the highest among all three methods. Therefore, our proposed method preserves prediction precision and offers a good balance between PI length and PI coverage.

\begin{remark}
We would like to compare the site-level with the study-level framework in depth from the models themselves. First, it needs to emphasize that the difference from data source should be considered. In fact, it is difficult to get access to site-level data in practice, so the number of underlying historical studies for the site-level model is often much less than that for the study-level model. Therefore, the 95\% PI from site-level model is wide. Second, the site-level model is more complex from methodology perspective. The site-level framework consists of three segments (i.e., country start-up, site activation and subject enrollment), which can fit for a broad-sense enrollment process. Due to the lack of studies with site-level data in practice, the numeric performance may not have been pushed to the extreme. As a result, practitioners should evaluate data availability before determining which model to use.
\end{remark}

\begin{table}[h]
\begin{tabular}{p{0.05\linewidth}|p{0.095\linewidth}p{0.095\linewidth}p{0.095\linewidth}p{0.095\linewidth}p{0.07\linewidth}p{0.07\linewidth}p{0.07\linewidth}p{0.07\linewidth}}
\hline
Study & Actual Enrollment  (mo) & Predicted Enrollment  (mo) & Prediction Error (mo) & 95\%PI (mo) & Within 95\%PI & Within +/-1 month & Within +/-2 month & Within +/-3 month \\ \hline
1     & 26.4                    & 20.5                       & -5.9                  & (13.7,31.3) & YES           & NO             & NO                & NO                \\
2     & 17.5                    & 20.2                       & 2.7                   & (13.5,35.4) & YES           & NO             & NO                & YES               \\
3     & 11.3                    & 10.1                       & -1.1                  & (6.4,17.3)  & YES           & NO             & YES               & YES               \\
4     & 10.0                    & 10.0                       & -0.0                  & (6.9,17.3)  & YES           & YES            & YES               & YES               \\
5     & 21.1                    & 22.1                       & 1.0                   & (15.7,36.7) & YES           & NO             & YES               & YES               \\
6     & 16.1                    & 35.0                       & 18.9                  & (16.2,Inf)  & YES           & NO             & NO                & NO                \\
7     & 10.8                    & 11.6                       & 0.8                   & (8.11,17.7) & YES           & YES            & YES               & YES               \\ \hline
\end{tabular}
\caption{\label{tab:table5} Performance of the Site-level model Zhong et al. \cite{zhong2022sitelevel}.}
\end{table}

\begin{table}[h]
\begin{tabular}{p{0.05\linewidth}|p{0.095\linewidth}p{0.095\linewidth}p{0.095\linewidth}p{0.095\linewidth}p{0.07\linewidth}p{0.07\linewidth}p{0.07\linewidth}p{0.07\linewidth}}
\hline
Study & Actual Enrollment  (mo) & Predicted Enrollment  (mo) & Prediction Error (mo) & 95\%PI (mo) & Within 95\%PI & Within +/-1 month & Within +/-2 month & Within +/-3 month \\ \hline
1     & 26.4                    & 17.7                       & -8.7                  & (16.9,18.4) & NO            & NO                & NO                & NO                \\
2     & 17.5                    & 13.5                       & -4.0                  & (12.7,14.2) & NO            & NO                & NO                & NO                \\
3     & 11.3                    & 9.1                        & -2.2                  & (8.5,9.9)   & NO            & NO                & NO                & YES               \\
4     & 10.0                    & 9.6                        & -0.4                  & (8.8,10.4)  & YES           & YES               & YES               & YES               \\
5     & 21.1                    & 15.3                       & -5.8                  & (14.5,16.1) & NO            & NO                & NO                & NO                \\
6     & 16.1                    & 31.1                       & 15.0                  & (27.4,35.0) & NO            & NO                & NO                & NO                \\
7     & 10.8                    & 11.7                       & 0.9                   & (10.8,12.7) & YES           & YES               & YES               & YES               \\ \hline
\end{tabular}
\caption{\label{tab:table6} Performance of the traditional model in Zhong et al.\cite{zhong2022sitelevel}.}
\end{table}

%% file: sections/discussion.tex
\label{sec:discussion}
For the problem of predicting study duration at early planning stage, the proposed study-level forecasting model is neat and accurate to characterize entire enrollment by the subject recruitment and the site initiation parts. Through numeric analysis, our model conducts a balanced forecast between PI length and prediction error, which essentially stand for the trade-off between variance and bias. 

In terms of determining the number of sites in each country, our framework assumes they are specified at the beginning. However, at early stage of study preparation, it is common to only have a list of countries to carry out the trial and a total number of sites to initiate. As a consequence, there exist a need to get more insights from historical data to guide the start-up of counties as well as sites. 
In our performance evaluation, we used the number of sites from a given list of countries and approximated the remaining sites by corresponding median estimates of model parameters based on all countries on the list.
Similarly, practitioners can use historical data to obtain a recommendation of the number of sites of each country based on, for instance, maximum of opened sites in each study, site opening rate and country start-up days.

With regard to the prediction of site initiation, we found the FSFD seemed to have a negative bias in our performance evaluation in Section~\ref{sec:model_performance}, though it was not conclusive from validation of only 7 studies. Despite that, the overall predictions of study duration were generally good. This phenomenon indicates a potential improvement on the site activation modelling. One possible remedy is to build a separate country start-up model in addition to site initiation as in Zhong et al. \cite{zhong2022sitelevel} to better model $t_i$. Another way is to consider zero-inflation Poisson model for $u_{i,j}$, where an additional underlying process is introduced to determine whether a count is zero or non-zero. Then, the underlying process brings more chance to zero observations while a large rate parameter will accompany to balance the overall modelling. 
Since it is out the scope of this project, we will leave the training of zero-inflation model to the future. 

For the subject accrual rate parameter, $\mu$, is set to be a common parameter across country in our model. A natural variation may be to estimate the rate along with country. However, as mentioned in the quasi-Poisson model set-up, historical data availability does not encourage an integrated structure incorporating too many factors like random effect and country effect. Hence, we would leave such modification as a humble remark here.